\documentstyle[11pt,newpasp,twoside,epsf]{article}
\def\ltsima{$\; \buildrel < \over \sim \;$}
\def\simlt{\lower.5ex\hbox{\ltsima}}
\def\gtsima{$\; \buildrel > \over \sim \;$}
\def\simgt{\lower.5ex\hbox{\gtsima}}

\def\ls{{_<\atop^{\sim}}}
\def\gs{{_>\atop^{\sim}}}

\def\edcomment#1{\iffalse\marginpar{\raggedright\sl#1\/}\else\relax\fi}
\reversemarginpar

\begin{document}
\title{Chandra and FUSE view of the WHIM: the Local Group and beyond}
\author{Fabrizio Nicastro}
\affil{Harvard-Smithsonian Center for Astrophysics, 60 Garden Street, MS-83, 
Cambridge, MA 02138, U.S.A.}

\begin{abstract}
In this contribution, I review the current observational 
evidence for the existence of filaments of Warm-Hot Intergalactic 
Medium (WHIM). In particular, I first focus on the controversial 
issue of the identification of the $z\sim 0$ highly ionized far-ultraviolet 
(i.e. OVI) and X-ray (i.e. OVII, OVIII and NeIX) absorbers with either 
a very tenuous and diffuse WHIM filament, or with much denser 
condensations of material at large distances in the Galactic halo. 
I then present our recent detection (confidence level $> 
3\sigma$) of the OVII WHIM at $z > 0$ and derive an estimate of the total 
number of baryons contained in this hard-to-detect phase of the IGM. 
\end{abstract}

\section{Introduction}
Despite recent progress in cosmology, very little is still known 
about the location of baryons in the local Universe. 
The extraordinary WMAP ({\em Wilkinson Microwave Anisotropy Probe}) 
measurements of the Cosmic Microwave Background (CMB) anisotropies favor a 
$\Lambda$-CDM scenario in which the baryon density in the Universe 
amounts to $\Omega_b = (4.4 \pm 0.4) h_{70}^{-2}$ \% of the total 
matter-energy density, with a baryon-to-dark-matter ratio of $0.17 \pm 0.1$ 
(Bennet et al., 2003). This number agrees well with predictions 
by the standard 'big bang nucleosynthesis' when combined with light element 
ratios ($\Omega_b = (3.9 \pm 0.5) h_{70}^{-2}$ \%: Burles \& Tytler, 1998), 
and also with the actual number of baryons at redshifts 
larger than 2 in the 'trees' of the Ly$\alpha$ Forest: 
$\Omega_b \ge 3.5 h_{70}^{-2}$ \% (Rauch, 1998; Weinberg et al., 
1997). This concordance cosmology represents a major advance. However, 
the number of baryons detected in the ``virialized Universe'' at $z \ls 2$ 
(i.e. stars, neutral hydrogen associated with galaxies, 
and X-ray emitting gas in clusters) is more than a factor of 2 smaller 
than the concordance value, $\Omega_b < 2 h_{70}^{-2}$ \% (2$\sigma$; e.g. 
Fukugita, Hogan, \& Peebles, 1998). 
It has also been known since 1959 (Kahn \& Woltjer, 1959) that, locally, 
more than $1.5 \times 10^{12}$ M$_{\odot}$ are needed to dynamically 
stabilize our own Local Group. 
So, one fundamental question arises: where have the baryons been hiding 
for the last $\sim 10$ Gyrs of life of the Universe?

\medskip
Hydrodynamical simulations for the formation of structures in the 
Universe, predict that in the present epoch ($z \ls 1-2$), 
half of the normal baryonic matter in the Universe (the ``missing baryons'') 
is in a tenuous (overdensities $\delta \simeq 5-50$, 
relative to the mean baryon density in the Universe)``warm phase'' (the so 
called 'Warm Hot Intergalactic Medium': WHIM), shock-heated to temperatures 
of $10^5-10^7$ K during the continuous process of structure formation (e.g. 
Hellsten et al., 1998; Cen, \& Ostriker, 1999; Dav\'e et al., 2001). 
These hot filaments are now detectable with current instruments. 

\section{The $z \sim 0$ Highly Ionized Absorbers: Local Group 
{\em versus} Galaxy Halo}. 

Two years ago we reported (Nicastro et al., 2002, hereinafter N02) 
the first detection of highly ionized O and Ne 
absorption [OVII$_{K\alpha}(\lambda21.602)$, OVIII$_{K\alpha}
(\lambda18.97)$, NeIX$_{K\alpha}(\lambda13.447)$] at redshift consistent 
with zero, in the {\em Chandra} Low Energy Transmission 
Grating (LETG) spectrum of the blazar PKS~2155-304 (Fig 1). 
%
\begin{figure}[t]
\epsfysize=2.4in 
\epsfxsize=2.4in 
\hspace{3cm}
\epsfbox{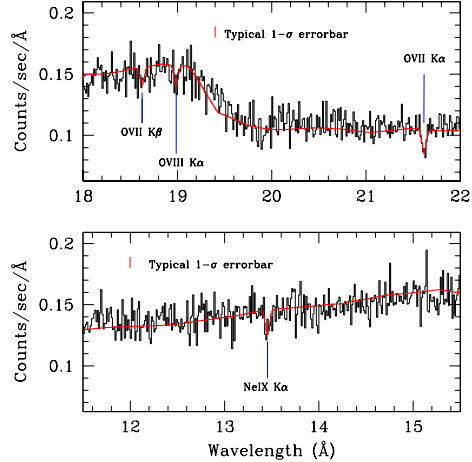} 
\vspace{0in}\caption[h]{\footnotesize Two portions of the {\em Chandra-LETG} 
spectrum of PKS~2155-304. OVII K$\alpha$ and NeIX K$\alpha$ resonant 
absorption lines at $z\sim 0$, from our Local Group WHIM filament, are 
clearly present at high significance, and associated OVII K$\beta$ and 
OVIII Ly$\alpha$ are detected at lower significance.}
\end{figure}
%
Detection of Li-like oxygen absorption [OVI$_{2s\rightarrow2p}(\lambda = 
1031.93, 1037.62)$] was also reported in the Far-Ultraviolet FUSE 
spectra of this (Sembach et al., 2000; Savage et al., 2000; N02) and many 
other lines of sight (Sembach et al., 2003: hereinafter S03; 
Savage et al., 2003; Nicastro et al., 2003: hereinafter N03). 
The resolution of the FUSE spectrometers is 
about 30 times better than that of the LETG, and clearly 
sufficient to resolve the OVI complex in at least two components: 
a high velocity (HV-OVI) component at $v_{LSR} \simeq -300$ to -100 km 
s$^{-1}$, and a narrower and more symmetric low velocity (LV-OVI) 
component at $v_{LSR} \simeq -30$ to +100 km s$^{-1}$ (see Fig. 3a in N02). 

\subsection{A Controversial Solution} 
Figure 2 shows that the UV absorbers (both, HV and LV) are incompatible 
with the X-ray absorbers in the hypothesis of purely collisional ionized 
gas with typical Interstellar Medium (ISM) densities (solid lines and 
superimposed 2$\sigma$ intervals for the ion abundance ratios). 
However, a single temperature solution (at T$\sim 6.3 \times 10^5$ K, 
dashed vertical line in Fig. 2) can be found for the HV-OVI and the 
X-ray absorbers if the volume density of the, mainly collisional ionized, 
gas is low enough (i.e. $n_e < 10^{-5}$ cm$^{-3}$) to allow photoionization 
by the extragalactic X-ray background to contribute (to $\sim 10$ \% at 
$n_b = 4 \times 10^{-6}$ cm$^{-3}$) to the overall ionization balance in 
the gas. 
This is because the relative population of He-like and H-like ions of O and 
Ne, compared to Li-like O, starts rising at relatively lower temperatures 
(i.e. T$\sim 3-7 \times 10^5$ K), compared to pure collisional ionization 
model (dashed lines and superimposed 2$\sigma$ intervals for the ion 
abundance ratios. See N02 for further details). 
%
\begin{figure}
\epsfysize=2.4in 
\epsfxsize=2.4in 
\hspace{1.3in} 
\epsfbox{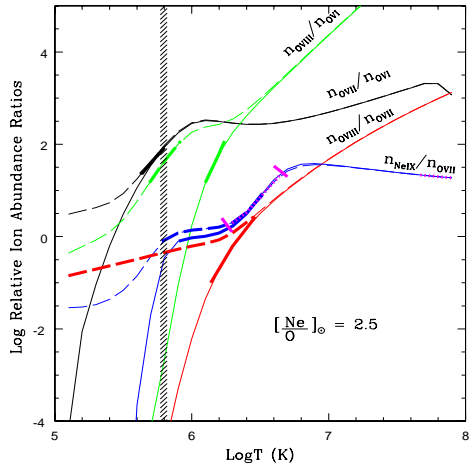}
\vspace{0in}\caption[h]{\footnotesize Expected ion abundance ratio curves 
for high-density gas in pure collisional ionization equilibrium (solid 
curve), and low density gas ($n_e \simeq 4-6 \times 10^{-6}$ cm$^{-3}$: 
dashed curves). Thick segments on these curves are allowed 2$\sigma$ 
intervals for the measured EW line ratios.}
\end{figure}
%
This solution implies a baryon volume density of $n_b = 4-6 \times 
10^{-6}$ cm$^{-3}$, corresponding to overdensities of $\delta = 20-30$ 
relatively to the average baryon density in the Universe. Assuming 
homogeneity, and an [H/O] ratio of 0.3 solar, this in turn gives a 
path-length of the absorber along this line of sight of $\sim 2-3$ Mpc, 
so locating the absorber in the intergalactic space surrounding the Galaxy. 
Assuming a transverse size of this filament of 1 Mpc, gives a baryonic 
mass of $\sim (0.6-2) \times 10^{12}$ M$_{\odot}$. 
Inhomogeneity would imply a smaller size of the absorber, and so 
a lower mass. If, for example,  the diffuse intergalactic medium would 
condensate when approaching the virialized structures of the Local Group, 
so mixing up with the ISM in the high galactic halo, up to densities of 
$n_b \sim 4 \times 10^{-4}$ cm$^{-3}$, and if this layer of denser gas 
provided, say, 75 \% of the total observed equivalent baryon column, 
then the total filament would be about 1 Mpc long along the line of sight 
to PKS~2155-304 (i.e. about the distance to M~31), with the dense 
shell contributing about 10 \% to the total mass of about $2 \times 
10^{11}$ M$_{\odot}$. 

The above solution (and consequent interpretation) is, of course, not unique. 
The alternative is that the X-ray and the ultraviolet 
absorbers are independently produced by a multiphase ISM (e.g. 
Heckman et al., 2002), or by both a hot 
multiphase ISM (the UV lines) and a diffuse local group medium 
or Galactic Corona (the bulk of the X-ray lines). 
The importance of this controversy lies in the total mass implied by the 
two different interpretations: in one case (Local Group WHIM: LG-WHIM) 
the implied mass is comparable, if not larger, than that of the most 
massive visible constituents of the Local Group, with strong implications 
for the Local Group dynamics and its stability, while in the Galactic 
scenario the gas mass would add up to a maximum of $\sim 1/100$ of the 
Galaxy's mass. 

\subsubsection{The Velocity Segregation of the HV-OVI} 
One of the main pieces of evidence supporting the extragalactic solution 
(i.e. LG-WHIM solution), is the segregation 
of the HV-OVI in the Local Standard of Rest (LSR) frame. 
About 85 \% of the lines of sight towards extragalactic sources 
observed by FUSE, show HV-OVI absorption (S03). 
Moreover, as demonstrated 
by N03, the majority of these absorbers appear to be at rest 
in the Local Group Standard of Rest (see Fig. 2 in N03). 
The velocity distribution of these absorbers is highly 
segregated in the Local Standard of Rest (LSR), with all negative-velocity 
absorbers lying at Galactic longitude $0^{\deg} \le l < 180^{\deg}$, and 
all positive-velocity absorbers lying in the opposite hemisphere (Fig. 2 
in N03). This particular distribution is expected for objects at rest in 
the intergalactic space surrounding our Galaxy, with the kinematics in the 
LSR being an effect of the rotation of the Sun in the Galaxy's plane. This 
suggestion is confirmed by the progressive reduction of the degree of 
segregation (and so of the amplitude of the average velocity vector) 
of the HV-OVI absorbers for translations of the velocity 
vectors into the Galactic Standard of Frame (GSF) and the Local 
Group Standard of Frame (LGSR). In the LGSR the HV-OVI distribution appears 
random, and the amplitude of the average velocity vector is close to the 
FUSE resolution
\footnote{We stress that, unlike the case of the HI High Velocity Clouds, 
for the HV-OVI the amplitudes of the average velocity vector in the GSR 
and LGSR are inconsistent with each other at $> 90$ \% confidence level.}
. N03 concluded that the HV-OVI absorbers are at rest in the Local 
Group and so trace either a very extended Galactic Corona (Galactocentric 
distances $\gs 150-200$ kpc) or a LG-WHIM filament (see also Wakker, 2003, 
these proceedings). 

S03 have argued against this interpretation. They 
noted that the above evidence is a necessary, but not sufficient condition 
to prove the LG-WHIM origin of the majority of the HV-OVI. While we 
certainly agree with this point, we want to stress here that their main 
argument against the LG-WHIM interpretation is the wide correspondence 
between some of the structures traced by the HV-OVI in the sky, and 
similar complexes seen in the hydrogen High Velocity Clouds (HVCs) 
distribution, and thought to be at distances of few 
kpc from the Galaxy center. However, both N03 and S03 were able to 
tentatively identify (based on very generous spatial and kinematics 
associations: up to tens of degrees in space and hundred km s$^{-1}$ 
in velocity) 
only about 18-20 \% of the HV-OVI with HVCs with known or estimated 
Galactocentric distance. These objects have not been considered in the 
statistical analysis presented by N03. For several other HV-OVI in 
their sample, S03 claim the association with known HVC structures (i.e. 
the Magellanic Stream, Complex C, Complex A), only on the base of 
the 'vicinity' (few degrees or tens of degrees apart) of these absorbers 
with the above HVCs structures. 

\subsubsection{Distribution of the Highly Ionized X-ray Absorbers} 
Incredibly high quality {\em Chandra} and/or {\em Newton}-XMM grating 
spectra are needed to detect X-ray absorption lines with typical 
WHIM intensities (i.e. EW($OVII_{K\alpha}$) $< 15$ m\AA). The LETG 
spectrum of the blazar PKS~2155-304 (N02) was collected with an exposure 
of 60 ks, and during a powerful source outburst ($F_{0.5-2 keV} = 
21$ mCrab $= 4.2 \times 10^{-10}$ erg s$^{-1}$ cm$^{-2}$). 
This guarranteed us about 700 counts per resolution element at the 
wavelength of the OVII$_{K\alpha}$ line, enough to detect absorption lines 
with EW$> 10$ m\AA, at a significance larger than 3$\sigma$. 
Unfortunately, however, the flux level of PKS~2155-304 during this 
observation, was about 10-20 times that of the brightest nearby 
extragalactic sources in their quiescent levels. Very long exposures 
are then normally needed to obtain a sizeable signal-to-noise selected 
sample of X-ray spectra of extragalactic 
sources, suitable to study the distribution of the highly 
ionized X-ray absorbers at $z\simeq 0$. To date, only 5 extragalactic 
sources have such good quality spectra: PKS~2155-304, Mkn~421, 3C~273, 
NGC~4593 and NGC~5548. All of them show OVII$_{K\alpha}$ absorption with 
similar intensities (i.e. about 10-20 m\AA: N02, Rassmussen et al., 2003, 
Fang et al., 2003). 

\subsubsection{Galactic HV-OVI or Ionized X-ray Absorbers Wanted} 
Despite efforts to look for Galactic (or 
between the Galaxy and the two Magellanic Clouds) HV-OVI or highly 
ionized X-ray absorbers, none has been found yet. 

We searched the literature for reports of, non-intrinsic, highly 
ionized (i.e. OVII, OVIII or NeIX) absorbers at LSR velocity consistent 
with $-500 \ls v_{LSR} \ls +500$ km s$^{-1}$, in the X-ray grating spectra 
of binaries either in our Galaxy or in the two Magellanic Clouds, 
and could not find any. 
We found about a dozen published {\em Chandra} and/or {\em Newton}-XMM 
observations of X-ray binaries. 
In few cases variable (i.e. intrinsic) highly ionized emission and/or 
absorption, have been detected (e.g. Ness et al., 2003, ApJ, 594, L127). 
The remaining do not show any high ionization absoprption line either 
intrinsic or due to intervening ISM. 

The situation is even clearer in the FUV. Zsarg\'o, et al., (2003) 
report no detection of HV-OVI against any of the 22 O-B stars 
observed by FUSE in the Galactic halo. They also note that ``if HV-OVI 
absorption was as common toward halo stars as toward extragalactic sources, 
one would expect 10-12 detections of HV-OVI absorption in'' their ``sample''. 
They further note that this evidence ``is very important because it justifies 
the association of the thick-disk OVI with the low-velocity gas''. While we 
definitely agree with this conclusion, we also think that the 
above evidence directly suggests the association of the HV-OVI with 
extragalactic gas. 
Howk, Sembach \& Savage (2003) also report no detection of HV-OVI in the 
first 10 kpc, along the line of sight to the globular cluster M~3. 
Finally, only Galactic (i.e. Low-Velocity) OVI absorption, or OVI absorption 
at the velocity of the two Magellanic Clouds, is detected against samples 
of O-B stars in the two Clouds (Danforth et al., 2002; Hoopes et al., 2002). 
The few exceptions have Galactic explanation (i.e. a Supernova Remnant in 
our Galaxy, along the line of sight to AV~229). 

\subsubsection{Lower Ionization High Velocity Absorption} 
One of the strongest arguments against the LG-WHIM interpretation 
for the HV-OVI and the X-ray absorbers, is the evidence that, along 
some lines of sight (i.e. PKS~2155-304, Mkn~509: Sembach et al., 
1999), the HV-OVI have lower ionization 
counterparts (Si II/III/IV and C II/III/IV), with similar centroid velocity. 
This is not compatible with an homogeneous, diffuse and very 
tenuous WHIM surrounding our Galaxy, and producing the X-ray and the HV-OVI 
absorption. One possibility is therefore that the different absorptions are 
produced by a multiphase ISM, with typical ISM densities. 
However, inhomogeneities in the LG-WHIM could also 
easily give rise to a multiphase medium. The most efficient source of 
cooling in a line emitting plasma at T$\sim 6 \times 10^5$ K, is the 
OVI$_{2s\rightarrow2p}$ doublet. This mechanism is quite 
inefficient at typical WHIM densities, giving e-folding cooling times 
of 80 Gyrs, i.e. larger than the Hubble Time. Such a LG-WHIM obviously 
could not have had time to cool down to temperatures typical of the 
observed low-ionization species. However, an increase by a factor of 100 in 
density in compressed layers in proximity of the virialized structures 
of the Local Group, would reduce the cooling time to about 800 Myrs, 
about 2/25 the Galaxy's age. So, these denser layers of LG-WHIM, 
would have had time to cool down and produce the observed lower ionization 
species. 
Moreover, in the Local Group scenario, the velocity of the absorber in the 
LSR, is mainly due to our rotational motion in the Galaxy. 
So, along a given line of sight, any 'layer' (either Mpc- or kpc-thick) would 
produce absorption lines with roughly the same velocity in the LSR. 
The profile, however, would be different, depending mainly on the size of 
the absorber along the line of sight. This seems to agree with the profiles 
observed in low and high-ionization lines. 

\section{The WHIM at $z > 0$: the Line of Sight to Mkn~421}
While evidence for the $z\sim 0$ filament is quite robust both in the 
FUV and in X-rays
\footnote{Our Galaxy is embedded in the LG-WHIM and so, assuming spherical 
symmetry, any line of sight probes roughly about half the system and so 
sees large columns. The random orientation of intervening filaments, 
instead, greatly reduces the chances of observing such high columns.} 
, until very recently there had been no clear detection 
in the X-rays of OVII-OVIII WHIM absorption at $z > 0$ (either associated 
with the few known intervening OVI - Savage, Tripp, \& Lu, 1998; Tripp, 
Savage, \& Jenkins, 2000; Tripp \& Savage, 2000; Jenkins et al., 2003 - 
or faint - i.e. non-damped - H Ly$\alpha$ systems in the FUV, or at 
independent redshifts). 
The deepest observation of one of the brightest quasars 
at $z\sim 0.5$ (the 500 ks LETG observation of H~1821+643) 
is sensitive only to EW$ \gs 15$ m\AA (at $\ge 3\sigma$ 
level), and only 2-3 low-significance ($\sim 2\sigma$) detections of 
such systems have actually been claimed (Mathur, Weinberg, \& Chen., 
2003) along that line of sight, consistently with expectations. 

This is mostly due to the intrinsic steepness of the 'LogN-Log(EW)' of 
the WHIM OVII-OVIII absorption lines, at relatively large EW (see, e.g., 
Fig. 4 in Hellsten et al., 1998). 
A single OVII K$\alpha$ system with $EWs \ge 18$ m\AA\ is expected along 
a random line of sight, up to $z = 0.5$. This number rises to 
8, for only twice weaker systems (EW$ \sim 9$ m\AA). 

\medskip
Very high signal to noise {\em Chandra}-LETG or {\em Newton}-XMM RGS 
spectra of extragalactic sources are therefore needed to detect the WHIM. 
These can be provided with either multi-megasecond observations of a single 
line of sight, or observing Blazars when they undergo an outburst phase. 
We decided to pursue this second strategy, and during the past year 
triggered two Target-of-Opportunity Observations (TOOs) of the Blazar Mkn 421 
($z = 0.03$, i.e. $d = 128$ Mpc, De Vaucouleurs, et al., 1991), 
during two exceptionally high flux 
outburst phases. The two observations were carried out with the {\em Chandra} 
ACIS-LETG (2002, October 26-27) and the HRC-LETG (2003, July, 1-2), 
lasted 100 ks each, and reported 0.5-2 keV fluxes of 100 and 40 mCrab, 
respectively. This allowed us to collect more than 10 million counts 
in the coadded first order spectrum, and more than 6000 counts per 
resolution element at 21.6 \AA. In turn, this guaranteed a 
3$\sigma$-sensitivity of $N_{OVII} \gs 10^{14.8}$ cm$^{-2}$, more than 
an order of magnitude fainter than the LG-WHIM. 
Between the two Chandra ToO observations, Mkn~421 has also been 
observed by FUSE, following our request of a TOO. 

Several absorption lines are clearly detected in the {\em Chandra} 
spectrum of Mkn~421 (Fig. 3a: left panel), from the relatively strong 
OVII K$\alpha$ line from the LG-WHIM at $\sim 21.6$ \AA\ down to the OVII 
K$\alpha$ line at the source redshift ($z = 0.03$). 
Our preliminary analysis suggests the identification of these 
lines with 3 different absorbers, one of which ($\lambda = 21.852$ \AA, 
$N_{OVII} = (1.2 \pm 0.3) \times 10^{15}$ cm$^{-2}$) a truly intervening 
WHIM system at $z = 0.0116$, close
\footnote{The H Ly$\alpha$ and the OVII$_{K\alpha}$ systems are shifted 
in velocity by $420 \pm 120$ km s$^{-1}$, which supports the existence of 
multiphase WHIM.} 
to the redshift of a known weak Hydrogen Ly$\alpha$ system (Shull, 
Stocke, \& Penton, 1996). 
Upper limits (at $3\sigma$) of $N_{OVIII} < 2 \times 10^{15}$ cm$^{-2}$ 
and $N_{OVI} < 2.2 \times 10^{13}$ cm$^{-2}$ are also set on the 
OVIII K$\alpha$ (Fig. 3b: right panels) and OVI$_{2s\rightarrow2p}$ 
transitions from this system. 
Based on these measurements we estimate 
a temperature of the intervening WHIM OVII filament of $T = (0.7-2.5) 
\times 10^6$ K, centered around the peak of the temperature distribution 
of the WHIM (e.g. Dav\'e et al., 2001). 

The number of OVII filaments per unit redshifts of $dN_{OVII}/dz 
\simeq 35$, derived from this single detection, is in good agreement 
with expectations from hydrodynamical simulations (Hellsten et al., 1998). 
This is at least twice as much as the corresponding number for OVI 
($dN_{OVI}/dz < 14$, Tripp, 2003, private communication), and so 
by far the largest reservoir of baryons in the local Universe.  
%
\begin{figure}
\epsfysize=2.7in 
\epsfxsize=2.7in 
\hspace{1in} 
\epsfbox{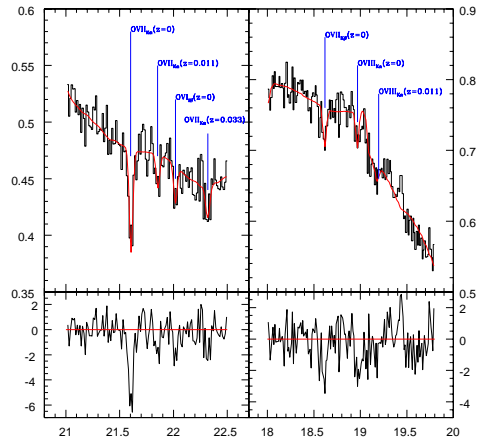}
\vspace{0in}\caption[h]{\footnotesize Two narrow portion of the coadded 
{\em Chandra} ACIS-LETG and HRC-LETG spectra of Mkn~421, centered around (a) 
the OVII$_{K\alpha}$ (left panels), and the OVIII$_{K\alpha}$ (right panels) 
transition. The solid continuous line in the bottom panels is the best 
fit polynomial plus absorption gaussians model to the data. Bottom panels 
show the residuals in $\sigma$ after the best fitting polynomial.}
\end{figure}
%

\begin{acknowledgements}
F.N. acknowledges support by the {\em Chandra} grant GO2-3122A. 
\end{acknowledgements}


\begin{references}
\reference{Bennett, C.L., et al., 2003, ApJS, 148, 1}
\reference{Burles, S., \& Tytler, D., 1998, Space Sci Rev, 84, 65}
\reference{Cen, R., \& Ostriker, J.P., 1999, ApJ, 514, 1}
\reference{Danforth, C.W., et al., 2002, ApJ, 586, 1179}
\reference{Dav\'e, R., et al., 2001, ApJ, 552, 473}
\reference{De Vaucouleurs, G., et al., 1991, ``Third Ref. Cat.  
of Bright Galaxies'', v.3.9}
\reference{Fang, T., Sembach, K.R. \& Canizares, C.R., 2003, ApJ, 586, L49}
\reference{Fukugita, M., Hogan, C.J., Peebles, P.J.E., 1998, ApJ, 503, 518}
\reference{Heckman, T.M., et al., 2002, ApJ, 579, 188}
\reference{Hellsten, U., et al., 1998, ApJ, 509, 56}
\reference{Hoopes, C.G., et al., 2002, ApJ, 569, 233}
\reference{Howk, J.C., Sembach, K.R., \& Savage, B.D., 2003, ApJ, 586, 249}
\reference{Jenkins, E.B., et al., 2003, AJ, 125, 2824}
\reference{Kahn, F.D., \& Woltjer, L., 1959, ApJ, 130, 705}
\reference{Mathur, S., Weinberg, D.H., Chen, X., 2003, ApJ, 582, 82}
\reference{Nicastro, F., et al., 2002, ApJ, 573, 157: N02}
\reference{Nicastro, F., et al., 2003, Nature, 421, 719: N03}
\reference{Rasmussen, A., et al., 2003, HEAD, 35.0201}
\reference{Rauch, M., 1998, ARA\&A, 36, 267}
\reference{Savage, B.D., Tripp, T.M., Lu, M., 1998, AJ, 115, 436}
\reference{Savage, B.D., et al., 2000, ApJ, 538, L27}
\reference{Savage, B.D., et al., 2003, ApJS, 146, 125}
\reference{Sembach, K.R., et al., 2000, ApJ, 538, L31}
\reference{Sembach, K.R., et al., 2003, ApJS, 146, 165: S03}
\reference{Shull, J.M., Stocke, J.T., \& Penton, S., 1996, AJ, 111, 72}
\reference{Tripp, T.M., Savage, B.D., \& Jenkins, E.B., 2000, ApJ, 534, L1}
\reference{Tripp, T.M., \& Savage, B.D., 2000, ApJ, 542, 42}
\reference{Weinberg, D., et al., 1997, ApJ, 490, 564}
\reference{Zsarg\'o, et al., 2003, ApJ, 586, 1019}

\end{references}
\end{document}